\documentclass[aps,manuscript,showpacs,showkeys,a4paper]{revtex4-1}
\usepackage{amsmath}
\usepackage{slashed,braket}
\usepackage{bm}
\usepackage{graphicx}
\usepackage{hyperref}

\begin{document}

\title{Meson Electro-Magnetic Form Factors in an Extended
  Nambu--Jona-Lasinio model including Heavy Quark Flavors}

\author{Yi-Long Luan}

\author{Xiao-Lin Chen}

\author{Wei-Zhen Deng}\email{dwz@pku.edu.cn}

\affiliation{School of Physics and State Key Laboratory of Nuclear
  Physics and Technology, Peking University, Beijing, 100871, China}

\begin{abstract}
  Based on an extended NJL model including heavy quark flavors, we
  calculate the form factors of pseudo-scalar and vector mesons.
  After taking into account the vector-meson-dominance effect, which
  introduces a form factor correction to the quark vector coupling
  vertices, the form factors and electric radii of $\pi^+$ and $K^+$ 
  pseudo-scalar mesons in the light flavor sector fit the experimental data
  well.  The magnetic moments of the light vector mesons $\rho^+$ and
  $K^{*+}$ are comparable with other theoretical calculations. The form
  factors in the light-heavy flavor sector are presented to compare
  with future experiments or other theoretical calculations.
\end{abstract}



\maketitle

\section{INTRODUCTION}

The Nambu--Jona-Lasinio (NJL) model \cite{Nambu:1961tp, Nambu:1961fr}
has been widely used in hadron physics as an effective model to study
chiral symmetry in the degree of quark freedom. Usually this model
deals with light hadrons composed of only light quark flavors
$u$, $d$, $s$ with $SU_f(3)$ symmetry \cite{Klevansky:1992qe,
  Klimt:1989pm, Vogl:1989ea, Vogl:1991qt}.

In hadron systems including heavy flavors, such as light-heavy mesons,
although the chiral symmetry is broken due to the mass of the heavy
quark, a complementary heavy flavor symmetry emerges and the so-called
heavy quark effective theory (HQET) was formulated for this using the
technique of $1/m_Q$ expansion \cite{Shifman:1986sm, Politzer:1988bs,
  Isgur:1989vq, Eichten:1989zv, Georgi:1990um, Grinstein:1990mj,
  Falk:1990yz, Mannel:1991mc}. In Ref.~\cite{Ebert:1994tv}, the NJL
model was extended to include heavy quark flavors to investigate such
light-heavy mesons like $D^{(*)}$ and $B^{(*)}$ mesons.

In our previous work \cite{Guo:2012tm}, we also tried to extend the NJL
model to include heavy flavors by expanding the NJL interaction
strengths in the inverse power of constituent quark masses according
to HQET. Based on this extension, we obtained the meson masses and
meson-quark coupling constants of all light and light-heavy mesons in
a unified way. Furthermore, the decay widths of the mesons were calculated
from those effective meson quark couplings \cite{Deng:2013uca}.

In this work, we further calculate the electromagnetic form factors of
mesons within this extended model.  Electromagnetic form factors play
an important role in our understanding of hadronic structure. The form
factors of the pseudo-scalar mesons $\pi$ and $K$ were measured in
several experiments \cite{Amendolia:1984nz, Amendolia:1986ui,
  Amendolia:1986wj} and in some previous theoretical works, the form
factors of $\pi$ and $K$ mesons were studied in the NJL model
\cite{Bernard:1988bx,Lutz:1990zc}. After considering the effect of
vector-meson-dominance of the vector mesons, such as the $\rho$ meson,
in the calculation, typically the form factors of $\pi$ fit the
experimental data well. Furthermore, the form factor of $\pi$ was also
studied in case of finite temperature with the NJL model
\cite{Schulze:1994fy}. Certainly the form factor of $\pi$ was studied
in many other theoretical approaches, such as the Dyson-Schwinger
equation using a confining quark propagator \cite{Roberts:1994hh},
light-cone or covariant quark wave functions
\cite{Chung:1988mu,Ito:1990ax}, and the lattice QCD method
\cite{Woloshyn:1985in,Nemoto:2003ng}. Also, with the QCD factorization
approach
\cite{Efremov:1979qk,Lepage:1979zb,Rothstein:2003wh,Bakulev:2004cu},
the form factor can be extrapolated to higher energy regions by taking
into account the perturbative QCD contribution.

The form factors of vector mesons have a rather more complicated structure.
Consequently they can provide us more information about vector mesons, such as
magnetic moments and quadrupole moments.  Presently there are only
theoretical results about the form factors of vector mesons.  Some
works have used the constituent quark model and the light front dynamics
\cite{Bakker:2002mt,Cardarelli:1994yq,Melo1997a} or Dyson-Schwinger equations
\cite{Bhagwat2008}. Lattice QCD calculation have been performed with the
three-point functions method \cite{Lee2008}, and the background
field method using only two-point functions
\cite{Hedditch:2007ex,Rubinstein:1995hc}. The magnetic moments of
vector mesons were also calculated by dynamics with the external
magnetic field \cite{Badalian2013}, and with QCD sum rules \cite{Aliev2009}.

There are a few papers studying the form factor of light-heavy mesons
\cite{Gomez-Rocha2012}. These focus on the electroweak form
factors. From the heavy flavor symmetry, those form factor should be
unifying described by the Isgur-Wise function when the heavy
flavor mass goes to infinity.

Here, we perform a systematic calculation of the meson form factors,
including pseudo-scalar mesons and vector mesons, of both the light
flavor sector and the light-heavy flavor sector, within the extended
NJL model. In the next section, we will introduce our model and
formalism. The numerical results and discussion will be presented in
Section~3.

\section{MODEL AND FORMALISM}

\subsection{Extended NJL model}

To deal with both light and heavy mesons in the Nambu-Jona-Lasinio
(NJL) model, in Ref.~\cite{Guo:2012tm} the four-fermion point
interactions are modified to
\begin{align}
  \mathcal{L}^F_4=G_V(\bar{q}\lambda^a_c\gamma^\mu q)
  (\bar{q}'\lambda^a_c\gamma^\mu q')
  +\frac{h}{m_qm_{q'}}[(\bar{q}'\lambda^a_c\gamma_\mu q')
    (\bar{q}\lambda^a_c\gamma_\mu q)
  +(\bar{q}\gamma_\mu\gamma_5\lambda^a_c q)
  (\bar{q}'\gamma_\mu\gamma_5\lambda^a_c q')]
\end{align}
where $\lambda^a$ are the generator of $SU(3)$ in color space and
$q,q'=u,d,s,c,b$ including both the light and the heavy flavors.  Here
the second part of the interaction is required to improve the spectra of
light vector mesons and the factor of $1/(m_qm_{q'})$ guarantees that
the symmetry of heavy flavors will still be hold in the heavy quark
limit according to HQET.

By solving the Bethe-Salpeter equation (BSE), we obtain the meson masses
and their coupling constants with quarks. We will use the effective
Lagrangian to describe the quark interaction in mesons. In the
case of $\pi$ and $\rho$, the effective Lagrangian reads
\begin{align}
  \mathcal{L}_{\pi qq}=&-g_{\pi q}\bar{q}i\gamma_5\bm{\tau}q\cdot\bm{\pi}
  -\frac{\tilde{g}_{\pi q}}{2m_q}\bar{q}\gamma_\mu\gamma_5\bm{\tau}q
  \cdot\partial^\mu\bm{\pi}, \\
  \mathcal{L}_{\rho qq}=&-g_{\rho q}\bar{q}\gamma_\mu\bm{\tau}q
  \cdot\bm{\rho}.
\end{align}
Here the couplings $g_{\pi q}$, $\tilde{g}_{\pi q}$ and $g_{\rho q}$ are
treated as constants since the energy of immediate quarks is
truncated to the low energy region in the NJL model.

In ref. \cite{Deng:2013uca}, we have calculated the strong and
radiative decays of vector mesons.  In this work, we will use the
above effective meson Lagrangian to further calculate the form factors of
mesons.

\subsection{Form factor of pseudo-scalar mesons}

The definition of the form factor of a pseudo-meson is given by
\begin{align}
\label{eq5}
\braket{\pi^+(p_2)|\bar{\psi}\gamma_\mu\psi|\pi^+(p_1)}=(p_1+p_2)_\mu F(q^2),
\end{align}
where $q=p_1-p_2$ is the transfer momentum. Its Feynman diagrams are
shown in Fig.~1
\begin{figure}
\begin{center}
\includegraphics[scale=0.7]{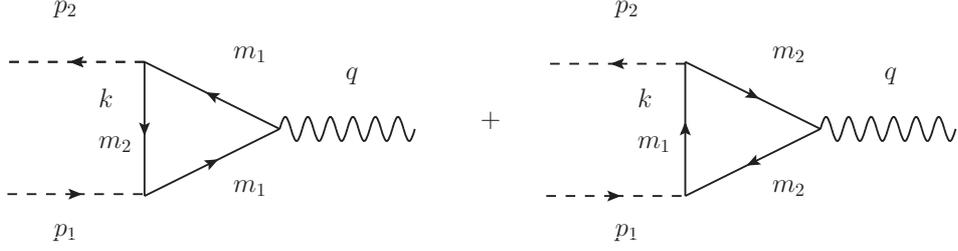}
\caption{Feynman diagrams of meson form factor.}
\end{center}
\end{figure}
where $m_1$ and $m_2$ are the masses of the constituent quarks in the
pseudo-scalar meson. Using the Feynman rules, the amplitude reads
\begin{align}
  (p_1+p_2)_\mu F(q^2)=&(p_1+p_2)_\mu [Q_1 F^{(1)}(q^2)+Q_2 F^{(2)}(q^2)] \\
  (p_1+p_2)_\mu F^{(1)}(q^2)=&-\text{Tr}\int\frac{d^4k}{(2\pi)^4} 
  \gamma_\mu S_1(k+p_1)
  i(g-\tilde{g}\frac{\slashed{p}_1}{m_1+m_2})i\gamma_5 S_2(k)
  \nonumber\\
  &\qquad\times i(g+\tilde{g}\frac{\slashed{p}_2}{m_1+m_2})
  i\gamma_5 S_1(k+p_2), \\
  (p_1+p_2)_\mu F^{(2)}(q^2)=&-\text{Tr}\int\frac{d^4k}{(2\pi)^4} 
  \gamma_\mu S_2(k-p_2)i(g+\tilde{g}\frac{\slashed{p}_2}{m_1+m_2})
  i\gamma_5 S_1(k) \nonumber\\
  &\qquad\times i(g-\tilde{g}\frac{\slashed{p}_1}{m_1+m_2})
  i\gamma_5 S_2(k-p_1),
\end{align}
where $F^{(1)}$ and $F^{(2)}$ are the form factors of the quark and
anti-quark respectively, $Q_i$ is the electron charge of $i$-th quark,
\begin{equation}
  S_i(p) = \frac{i}{\slashed{p}-m_i+i\epsilon}
\end{equation}
is the propagator of the $i$-th quark, and $g$ and $\tilde{g}$ are the
coupling constants of the pseudo-scalar meson obtained in our previous
work \cite{Guo:2012tm}.

In the Breit frame, $p_2^0-p_1^0 = 0$ and $\bm{p}_1=-\bm{p}_2$. We introduce
\begin{align}
  p_1=& p+q/2, & p_2=& p-q/2,
\end{align}
where $p\equiv \frac12(p_1+p_2) = (p_1^0,0)$, $q=(0,\bm{q})$.
Taking the direction of the $z$-axis along momentum $\bm{p}_1$, we find 
\begin{align}
  F^{(1)}(q^2) =& in_cn_f\int\frac{d^4k}{(2\pi)^4}
  \Big[g^2\frac{S_1}D-\frac{g\tilde{g}}{m_1+m_2}\frac{S_2}D
  -\frac{g\tilde{g}}{m_1+m_2}\frac{S_3}D
  +\frac{\tilde{g}^2}{(m_1+m_2)^2}\frac{S_4}D) \Big] \\
  F^{(2)}(q^2) =& F^{(1)}\Big(m_1\leftrightarrow m_2,q\to -q, p\to -p \Big),
\end{align}
where
\begin{align}
  S_1=& S_0(m_1,m_1,m_2), \\
  S_2=& m_1S_0(m_1,(k+p-q/2)^2/m_1,m_2)+m_2S_0(m_1,m_1,k^2/m_2), \\
  S_3=& m_1S_0((k+p+q/2)^2/m_1,m_1,m_2)+m_2S_0(m_1,m_1,k^2/m_2), \\
  S_4=& m_1^2 S_0((k+p+q/2)^2/m_1,(k+p-q/2)^2/m_1,m_2)+k^2S_0(m_1,m_1,m_2) \notag\\
  &+m_1m_2 S_0((k+p+q/2)^2/m_1,m_1,k^2/m_2)\notag\\
  &+m_1m_2S_0(m_1,(k+p-q/2)^2/m_1,k^2/m_2)\\
  D=&[(k+p+q/2)^2-m_1^2+i\epsilon)[(k+p-q/2)^2-m_1^2+i\epsilon]
  (k^2-m_2^2+i\epsilon),
\end{align}
and
\begin{align}
  S_{0}(m_1,m_2,m_3)\equiv&2[m_3(m_1+m_2)-2k\cdot(k+p)] \notag \\
  &+2 \frac{k\cdot p}{p^2} [m_3(m_1+m_2)-m_1m_2
  +p^2-k^2-q^2/4].
\end{align}
Note that the denominator $D$ of the integrand is invariant under the
transformation $\bm{k}\to -\bm{k}$.

The electromagnetic radius will be further obtained from the
derivative of the form factor via
\begin{equation}
  \label{eq2}
  r=\left[6\frac{dF}{dq^2}\right]^{1/2}_{q^2=0}.
\end{equation}
We have
\begin{equation}
  \label{r-sum-quark}
  r=\sqrt{Q_1 r_1^2 + Q_2 r_2^2},
\end{equation}
where
\begin{equation}
  r_i=\left[6\frac{dF^{(i)}}{dq^2}\right]^{1/2}_{q^2=0},
\end{equation}
is the radius of the $i$-th quark.

\subsection{Form factor of vector mesons}

The definition of the form factor of the vector meson reads
\cite{Brodsky:1992px,Arnold:1980zj}
\begin{align}
  \label{def-vec-form}
  \braket{\rho^+(p_2,\lambda_2)|\bar{\psi}\gamma_\mu\psi|\rho^+(p_1,\lambda_1)}
  =&-\epsilon^{\ast}(p_2,\lambda_2)\cdot\epsilon(p_1,\lambda_1)(p_1+p_2)_\mu 
  F_1(q^2)\nonumber\\
  &+[\epsilon_\mu(p_1,\lambda_1) q\cdot\epsilon^{\ast}(p_2,\lambda_2)
  -\epsilon^{\ast}_\mu(p_2,\lambda_2) q\cdot\epsilon(p_1,\lambda_1)]
  F_2(q^2)\nonumber\\
  &+\frac{q\cdot\epsilon^{\ast}(p_2,\lambda_2) 
    q\cdot\epsilon(p_1,\lambda_1)}{2m^2}
  (p_1+p_2)_\mu F_3(q^2),
\end{align}
where $\epsilon(p_1)$ and $\epsilon(p_2)$ are the polarization vectors
of the initial and the final vector meson respectively.  Based on the
Feynman diagrams, the LHS of Eq.~(\ref{def-vec-form}) can be written as
\begin{equation}
  \epsilon_\nu(p_1,\lambda_1)\epsilon^*_\lambda(p_2,\lambda_2) G_\mu^{\nu\lambda},
\end{equation}
where
\begin{align}
  G_\mu^{\nu\lambda}=& Q_1G_\mu^{(1)\nu\lambda} + Q_2G_\mu^{(2)\nu\lambda}, \\
  G_\mu^{(1)\nu\lambda}=&-\text{Tr}\int\frac{d^4k}{(2\pi)^4}
  \gamma_\mu S_1(k+p_1)ig_V \gamma^\nu S_2(k)ig_V\gamma^\lambda S_1(k+p_2), \\
  G_\mu^{(2)\nu\lambda}=&-\text{Tr}\int\frac{d^4k}{(2\pi)^4}
  \gamma_\mu S_2(k-p_2)ig_V \gamma^\lambda S_1(k)ig_V\gamma^\nu S_2(k-p_1) .
\end{align}

Still working in the Breit frame and taking the $z$-axis along the momentum
$\bm{p}_1$, the polarization vectors are chosen to be
\begin{align}
  \label{eq1}
  \epsilon(p_1,\pm)=&\frac1{\sqrt2} (0,1,\pm i,0), 
  &\epsilon(p_1,0)=&\frac1m (p_{1z},0,0,p_{10}), \nonumber\\
  \epsilon(p_2,\pm)=&\frac1{\sqrt2} (0,1,\mp i,0,)
  &\epsilon(p_2,0)=&\frac1m (p_{2z},0,0,p_{20}).
\end{align}
To retrieve $F_1$, we take the time component in
Eq.~(\ref{def-vec-form}) and find that
\begin{align}
  \epsilon_\nu(p_1,\lambda_1)\epsilon^*_\lambda(p_2,\lambda_2) G_0^{\nu\lambda}
  =&-\epsilon^{\ast}(p_2,\lambda_2)\cdot\epsilon(p_1,\lambda_1)(p_1+p_2)_0 
  F_1(q^2)\nonumber\\
  &+\frac{q\cdot\epsilon^{\ast}(p_2,\lambda_2) 
    q\cdot\epsilon(p_1,\lambda_1)}{2m^2}
  (p_1+p_2)_0 F_3(q^2).
\end{align}
Then $F_1$ can be obtained via the transverse polarization 
\begin{equation}
  \epsilon_\nu(p_1,\pm)\epsilon^*_\lambda(p_2,\pm) G_0^{\nu\lambda}
  =-\epsilon^{\ast}(p_2,\pm)\cdot\epsilon(p_1,\pm)(p_1+p_2)_0 
  F_1(q^2).
\end{equation}
To retrieve $F_2$, we take the spatial components in eq.~(\ref{def-vec-form})
and find that
\begin{align}
  \epsilon_\mu(p_1,\lambda_1)\epsilon^*_\nu(p_2,\lambda_2) G_i^{\mu\nu}=&
  -[\epsilon_i(p_1,\lambda_1) \bm{q}\cdot\bm{\epsilon}^{\ast}(p_2,\lambda_2)
  -\epsilon^{\ast}_i(p_2,\lambda_2) \bm{q}\cdot\bm{\epsilon}(p_1,\lambda_1)]
  F_2(q^2) \notag\\
  =&\left\{
    [\bm{\epsilon}(p_1,\lambda_1) \times\bm{\epsilon}^{\ast}(p_2,\lambda_2)]
    \times \bm{q} \right\}_i F_2(q^2) .
\end{align}
Still each form factor $F_j$ is a charge weight average of form
factors of quark and anti-quark in the vector meson,
\begin{equation}
  F_j(q^2)= Q_1 F_j^{(1)}(q^2) + Q_2 F_j^{(2)}(q^2),
\end{equation}
and
\[
F_j^{(2)}(q^2) = F_j^{(1)}\Big(m_1\leftrightarrow m_2,q\to -q, p\to -p \Big).
\]
Explicitly we obtain
\begin{align}
  F_1^{(1)}(q^2) =& \epsilon_\nu(p_1,+)\epsilon^*_\lambda(p_2,-) 
  G_0^{(1)\nu\lambda} \big/ (2p_0)\nonumber\\
  =& in_cn_fg_V^2\int\frac{d^4k}{(2\pi)^4}\frac{G_1}{D},
\end{align}
where
\begin{align}
  G_1=&4[(k+p)\cdot k - m_1m_2 + k_x^2 + k_y^2] \notag\\
  &-2\frac{p\cdot k}{p^2}[p^2-k^2-q^2/4-m_1^2+2m_1m_2 -2(k_x^2+k_y^2)],
\end{align}
and
\begin{align}
  F_2^{(1)}(q^2)=& -\frac{m_V}{p_0|\bm{q}|}
  \left[\frac{1-i}{\sqrt2}\epsilon_\mu(p_1,+)\epsilon^*_\nu(p_2,0) G_1^{(1)\mu\nu}
    +\frac{1+i}{\sqrt2}\epsilon_\mu(p_1,-)\epsilon^*_\nu(p_2,0) G_2^{(1)\mu\nu}
  \right]\nonumber\\
  =&in_cn_fg_V^2\int\frac{d^4k}{(2\pi)^4}\frac{G_2}{D},
\end{align}
where
\begin{align}
  G_2=&4[k\cdot(k+p)-m_1m_2 + k_z^2] \notag\\
  &-2\frac{k\cdot p}{p^2}[(k+p)^2-q^2/4-m_1^2+2(k_x^2+k_y^2)].
\end{align}
We will not consider the form factor $F_3$ in this work.

\subsection{Vector-Meson-Dominance and Quark Loop Correction}

According to the vector-meson-dominance picture, the $\pi$ and $K$ form
factor are dominated by the $\rho$, $\omega$ and $\phi$ intermediate
vector meson states \cite{Bernard:1988bx}. In the NJL model, the
vector-meson-dominance is represented by the correction to
quark-photon vertex as shown in the Feynman diagram in Fig.~2.
\begin{figure}
\begin{center}
\includegraphics[scale=0.8]{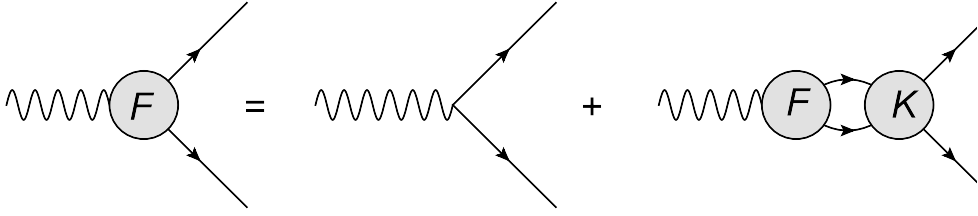}
\caption{Feynmann diagram of the loop correction}
\end{center}
\end{figure}
The correction will introduce a form factor to the constituent quark
\cite{Vogl:1989ea}. For the $i$-th quark
\begin{align}
F_q^{(i)}(q^2)=\frac1{1-K^VJ_{VV}^{(T)}},
\end{align}
where $K^V$ is the NJL vector coupling constant and $J_{VV}^{(T)}$
represents the transverse vector loop integral
\cite{Klimt:1989pm,Guo:2012tm}.  The meson form factor will be
modified to
\begin{equation}
  \label{F-sum-quark}
  F(q^2) = Q_1 F^{(1)}(q^2) F_q^{(1)}(q^2) + Q_2 F^{(2)}(q^2) F_q^{(2)}(q^2) .
\end{equation}

\section{NUMERICAL RESULTS}

The parameters of the extended NJL model were fixed by fitting the
meson mass spectra and decay constants in a previous work
\cite{Guo:2012tm}. The input parameters were the current masses of
light quarks and the constituent masses of heavy quarks, two
coupling constants and the 3-dimensional cutoff:
\begin{align}
&m_{u/d}^0=2.79\text{MeV},\qquad m_s^0=72.0\text{MeV},\nonumber\\
&m_c=1.62\text{GeV},\qquad\quad m_b=4.94\text{GeV},\nonumber\\
&\Lambda=0.8\text{GeV},\qquad\qquad G_V=2.41,\nonumber\\
&h=0.65.
\end{align}

Due to charge conservation, the form factor should be normalized
to $F(q^2=0)=1$ for any hadron carrying $+1$ charge.  We will make a
self-consistent calculation by using the theoretical values of meson
masses and the quark coupling constants together. This will guarantee
the strict normalization of the form factor at $q^2=0$
\cite{Schulze:1994fy}.  The theoretical values of pseudo-scalar and
vector mesons are listed in Table~\ref{tab1} and Table~\ref{tab2}
respectively.
\begin{table}
  \caption{The masses and quark coupling constants of pseudo-scalar mesons }
  \label{tab1}
\begin{ruledtabular}
\begin{tabular}{cccccc}
&$\pi$&$K$&$D$&$D_s$&$B$\\
\hline
mass(MeV)&139&496&1870&1940&5280\\
$g$&4.25&4.32&4.71&5.03&5.92\\
$\tilde{g}$&1.56&1.61&2.04&2.09&2.84
\end{tabular}
\end{ruledtabular}
\end{table}
\begin{table}
\caption{The masses and quark coupling constants of vector mesons}
\label{tab2}
\begin{ruledtabular}
\begin{tabular}{cccccc}
&$\rho$&$K^\ast$&$D^\ast$&$D_s^\ast$&$B^\ast$\\
\hline
mass(MeV)&771&918&1990&2120&5310\\
$g$&1.29&1.31&1.64&1.83&2.51
\end{tabular}
\end{ruledtabular}
\end{table}

\subsection{Pseudo-scalar mesons}

\begin{figure}
\begin{center}
\includegraphics[scale=0.72]{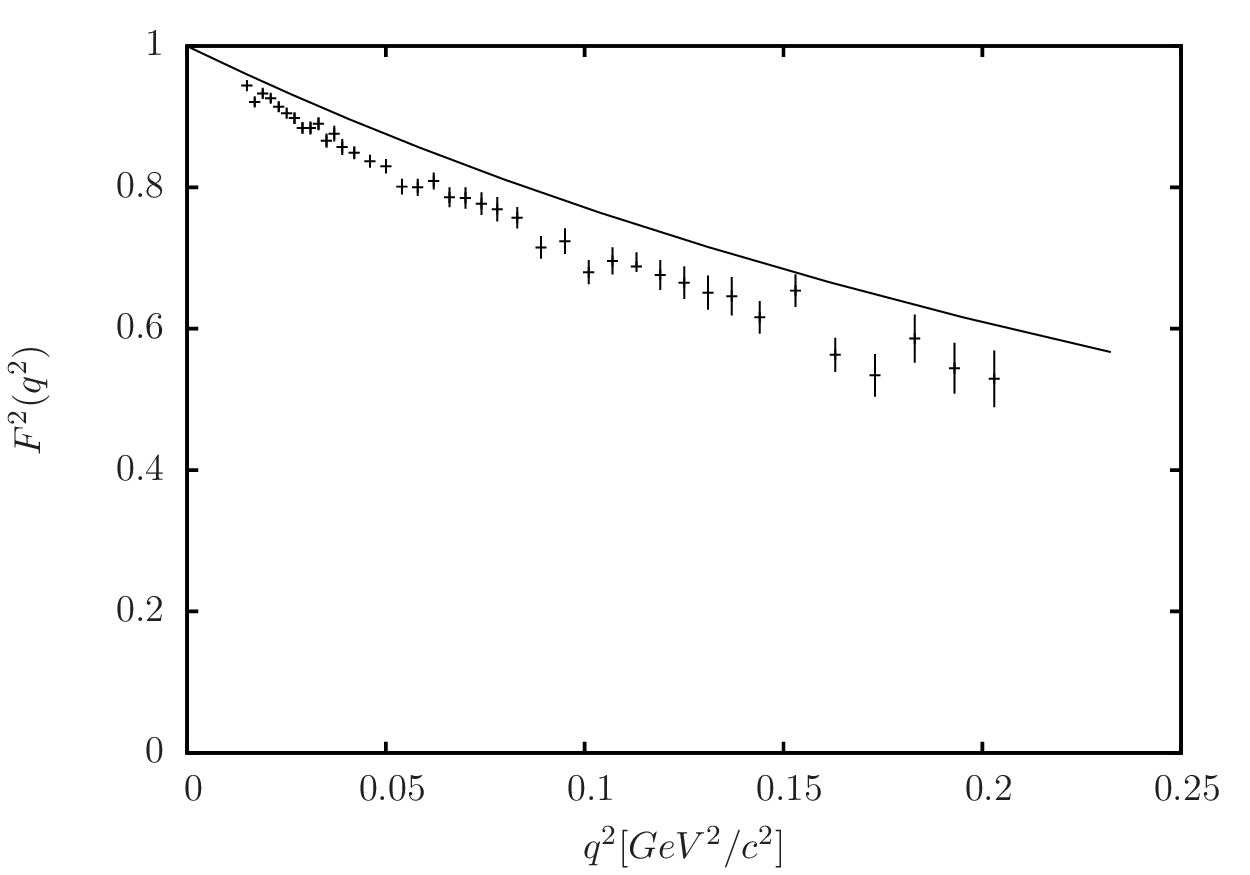}
\caption{The form factor of $\pi^+$ compared to the experimental
  data from Ref.~\cite{Amendolia:1986wj}.}\label{fig3}
\end{center}
\end{figure}

\begin{figure}
\begin{center}
\includegraphics[scale=0.72]{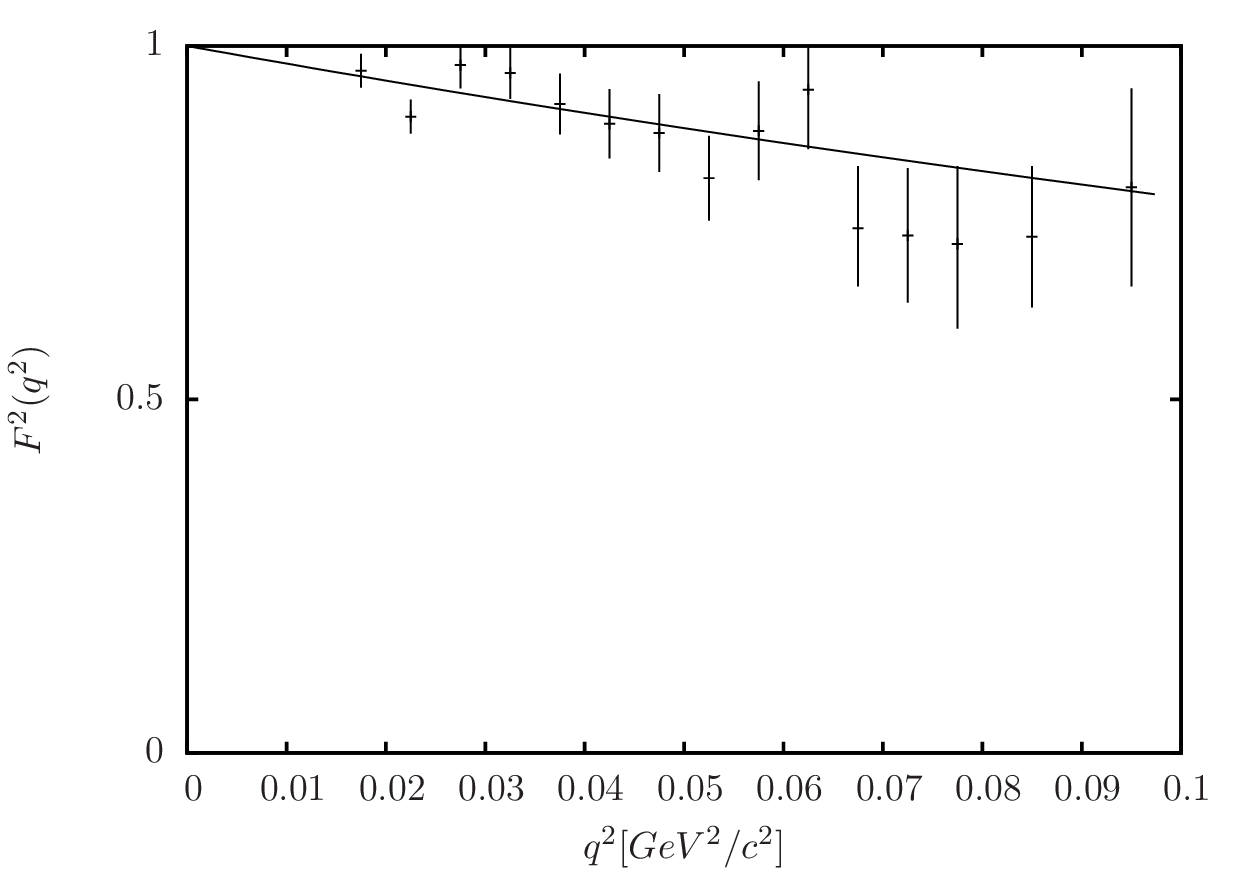}
\caption{The form factor of $K^+$ compared to the experimental data
  from Ref.~\cite{Amendolia:1986ui}.}\label{fig4}
\end{center}
\end{figure}

The form factors of $\pi^+$ and $K^+$ are compared with experimental
data in Fig.~\ref{fig3} and Fig.~\ref{fig4} respectively.
The theoretical results fit the experimental data well.  In the
theoretical calculation, the quark loop correction is included to account
for the important effect of vector-meson-dominance. 

\begin{figure}
\begin{center}
\includegraphics[scale=0.9]{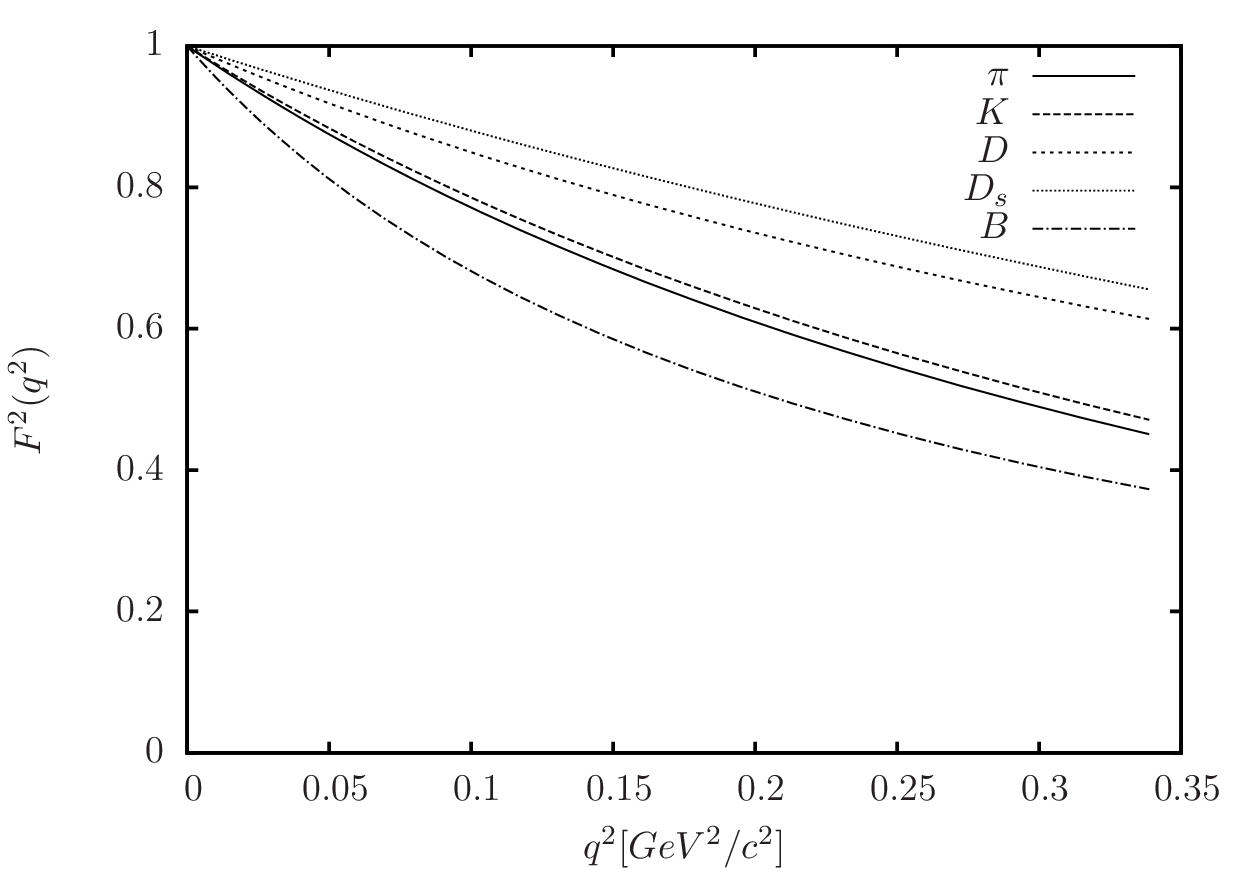}
\caption{The form factors of all positive pseudo-scalar mesons.}\label{fig6}
\end{center}
\end{figure}

The heavy-light pseudo-scalar mesons like $D^+$, $D^+_s$ and $B^+$ have
no experimental data for form factor yet. In Fig.~\ref{fig6}, we present
the form factors of all positive pseudo-scalar mesons.

The form factor at low momentum $q^2$ can be well
illustrated by the electromagnetic radius.  The radii of all
positive pseudo-scalar mesons are listed in Table~\ref{tab3}.
\begin{table}
  \caption{Comparison of electric radii $\braket{r^2}^{1/2}$ of 
    pseudo-scalar mesons 
    vs. experimental data. $r_1$  and $r_2$ are constituent quark 
    and anti-quark radii in the meson.}\label{tab3}
\begin{ruledtabular}
\begin{tabular}{ccccc}
  Meson&$\braket{r^2_1}(\text{fm}^2)$
  &$\braket{r^2_2}(\text{fm}^2)$&$\braket{r^2}^{1/2}_{\text{cal.}}(\text{fm})$&
  $\braket{r^2}^{1/2}_{\text{exp.}}(\text{fm})$
  \cite{Amendolia:1986ui,Amendolia:1986wj}\\
  \hline
  $\pi^+$&0.322&0.322&0.57&0.66\\
  $K^+$&0.341&0.206&0.54&0.56\\
  $D^{+}$&0.080&0.473&0.46&\\
  $D_s^{+}$&0.083&0.283&0.39&\\
  $B^{+}$&0.795&0.041&0.74&\\
\end{tabular}
\end{ruledtabular}
\end{table}

\begin{figure}
\begin{center}
\includegraphics[scale=0.85]{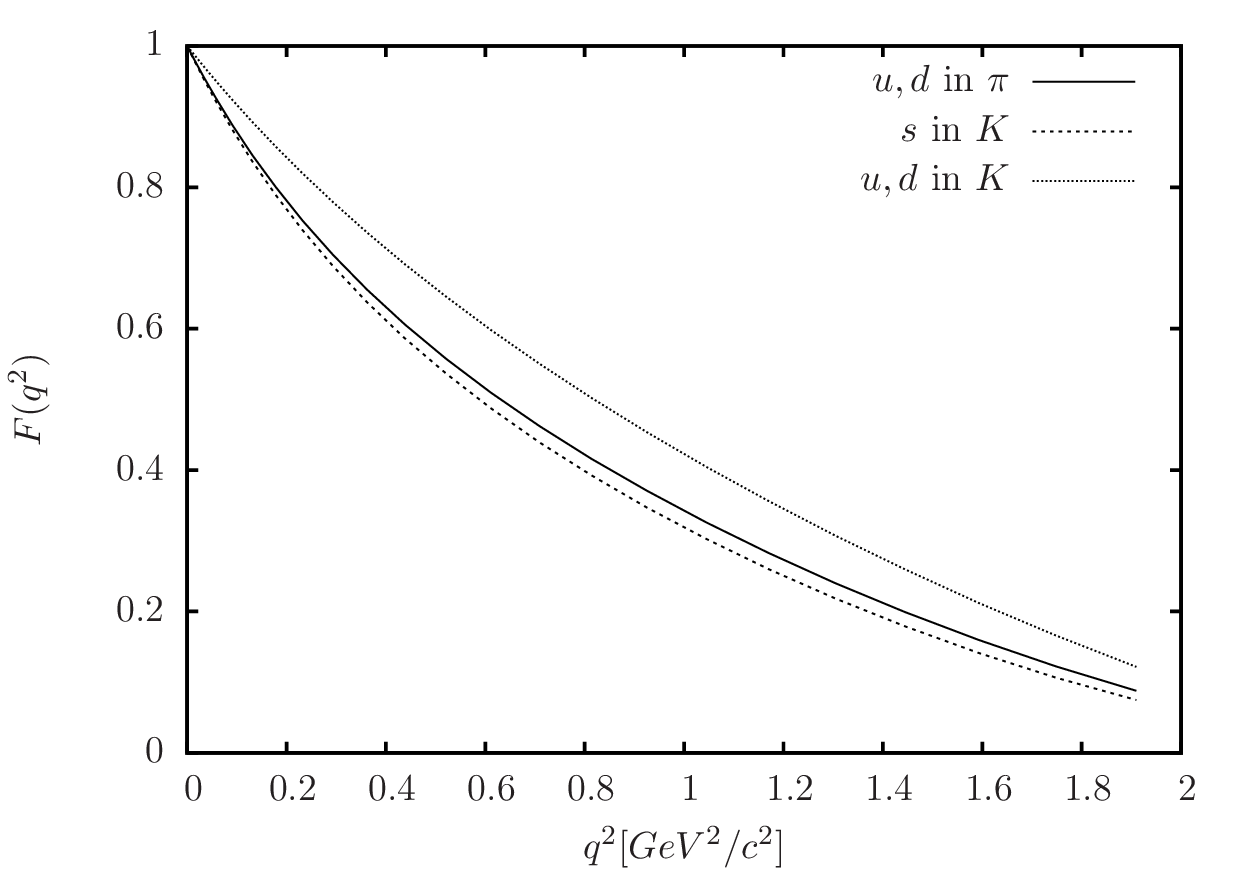}
\caption{The form factors of constituent quark and anti-quark in
  $\pi$ and $K$.}\label{fig5}
\end{center}
\end{figure}

As had been observed in Ref.~\cite{Bernard:1988bx}, it is the quark
loop correction of vector-meson-dominance that makes the $\pi$ radius bigger
than that of the $K$. In Eq.~(\ref{r-sum-quark}), the radius of a meson is
a charge weight average of individual quark radii. From
Eq.~(\ref{F-sum-quark}), after considering the quark loop correction, we
have
\begin{equation}
  \braket{r^2_i}=\braket{r_i^2}_{\text{int}}+\braket{r_i^2}_q,
\end{equation}
where
\begin{align}
  \braket{r_i^2}_{\text{int}}=&\left[6\frac{dF^{(i)}}{dq^2}\right]_{q^2=0}, \\
  \braket{r_i^2}_q=&\left[6\frac{dF^{(i)}_q}{dq^2}\right]_{q^2=0},   
\end{align}
are the ``intrinsic'' charge radius and the quark loop correction
respectively.  The quark loop correction decreases as the quark mass
increases, so the lighter quark has a larger radius than its heavier
partner in any meson. We show the individual form factors of quark and
anti-quark in $\pi$ and $K$ mesons in Fig.~\ref{fig5} and also list
the individual quark radii in Table~\ref{tab3}.

\begin{figure}
\begin{center}
\includegraphics[scale=0.9]{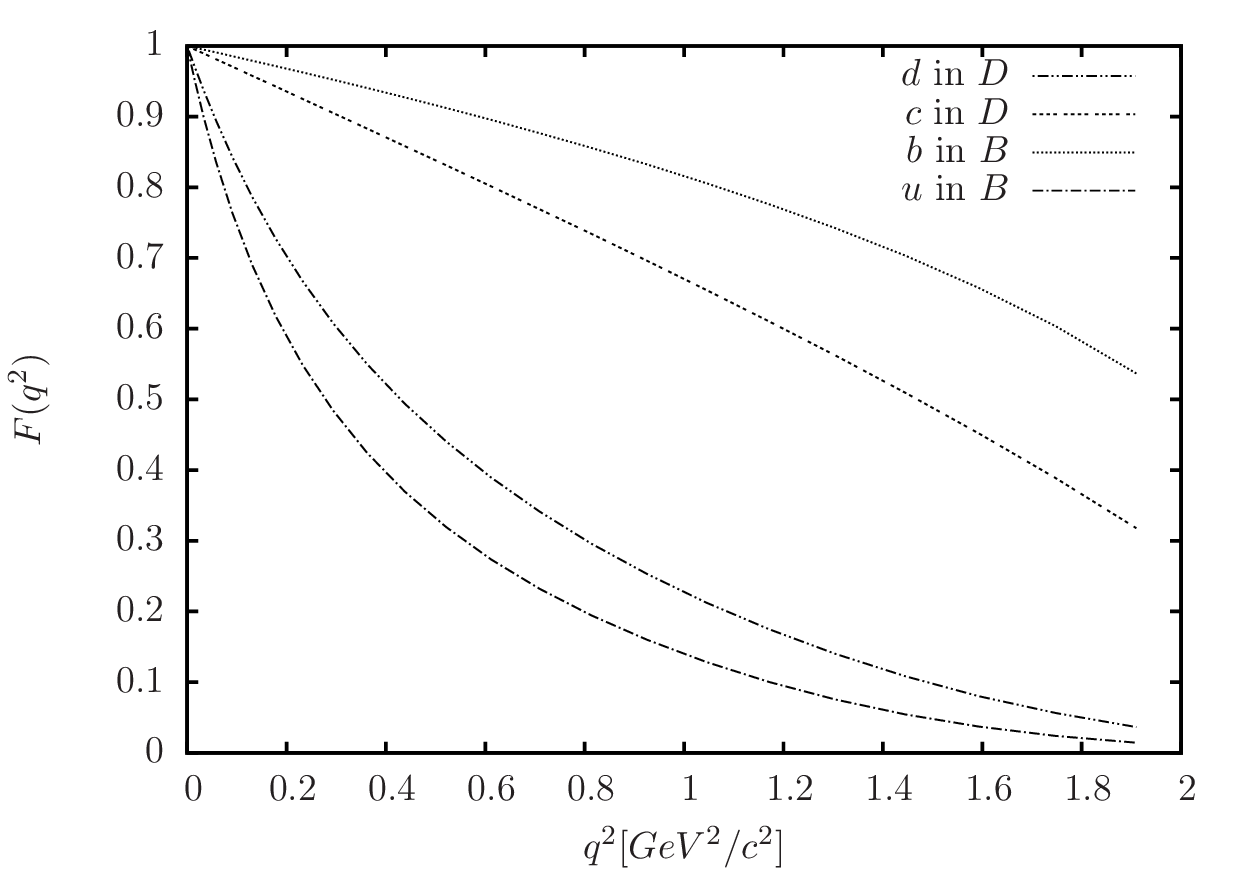}
\caption{The form factors of constituent quark and anti-quark in $B$ and
  $D$.}\label{fig7}
\end{center}
\end{figure}

Just like the $\pi^+$ and $K^+$, the radii of the light-heavy mesons
$\braket{r^2_{D^+}}^{1/2}$, $\braket{r^2_{D_s^+}}^{1/2}$ decrease as
the meson mass increases.  However the radius of the $B^+$ meson
increases by roughly a factor $2$.  This mainly because, in a
light-heavy meson, the heavy quark's contribution is much smaller than
that of the light one. If we neglect the contribution of the heavy
quark, in the $B^+$ meson, the $u$-quark has a $2/3$ charge weight of
contribution. The $d$-quark, on the other hand, has only a $1/3$
charge weight in the $D^+$ and $D^+_s$ mesons. The form factors of
individual constituent quarks in $D$ and $B$ are shown in
Fig.~\ref{fig7}.

\subsection{Vector mesons}

\begin{figure}
\begin{center}
\includegraphics[scale=0.9]{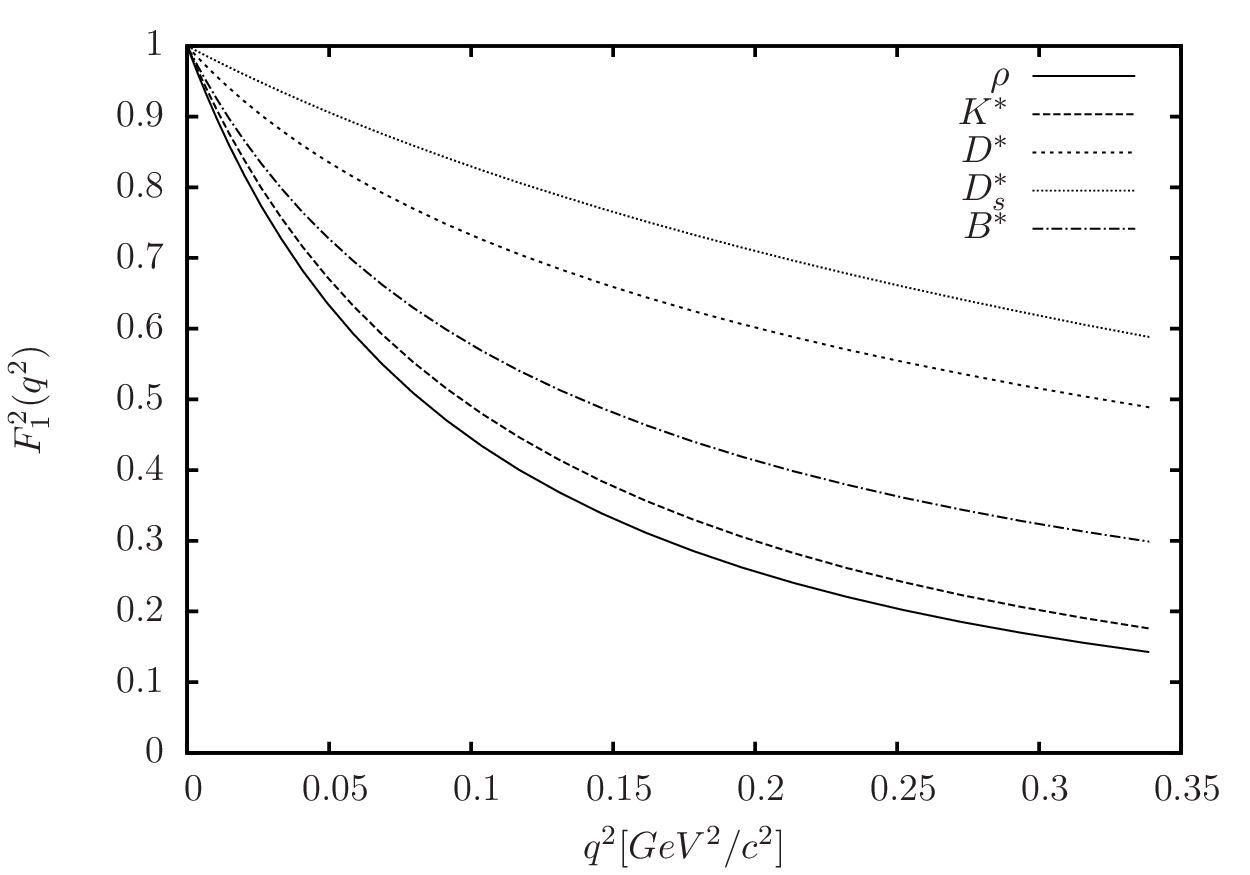}
\caption{$F_1$ of all positive vector mesons with respect to $q^2$}\label{fig8}
\end{center}
\end{figure}

The electric form factors $F_1$ of vector mesons are shown in
Fig.~\ref{fig8}.  The electric radii are listed in
Table~\ref{tab4}. Because all vector meson masses are close to
their thresholds, their bound energies are small and their radii are
larger than their pseudo-scalar partners.

\begin{figure}
\begin{center}
\includegraphics[scale=0.9]{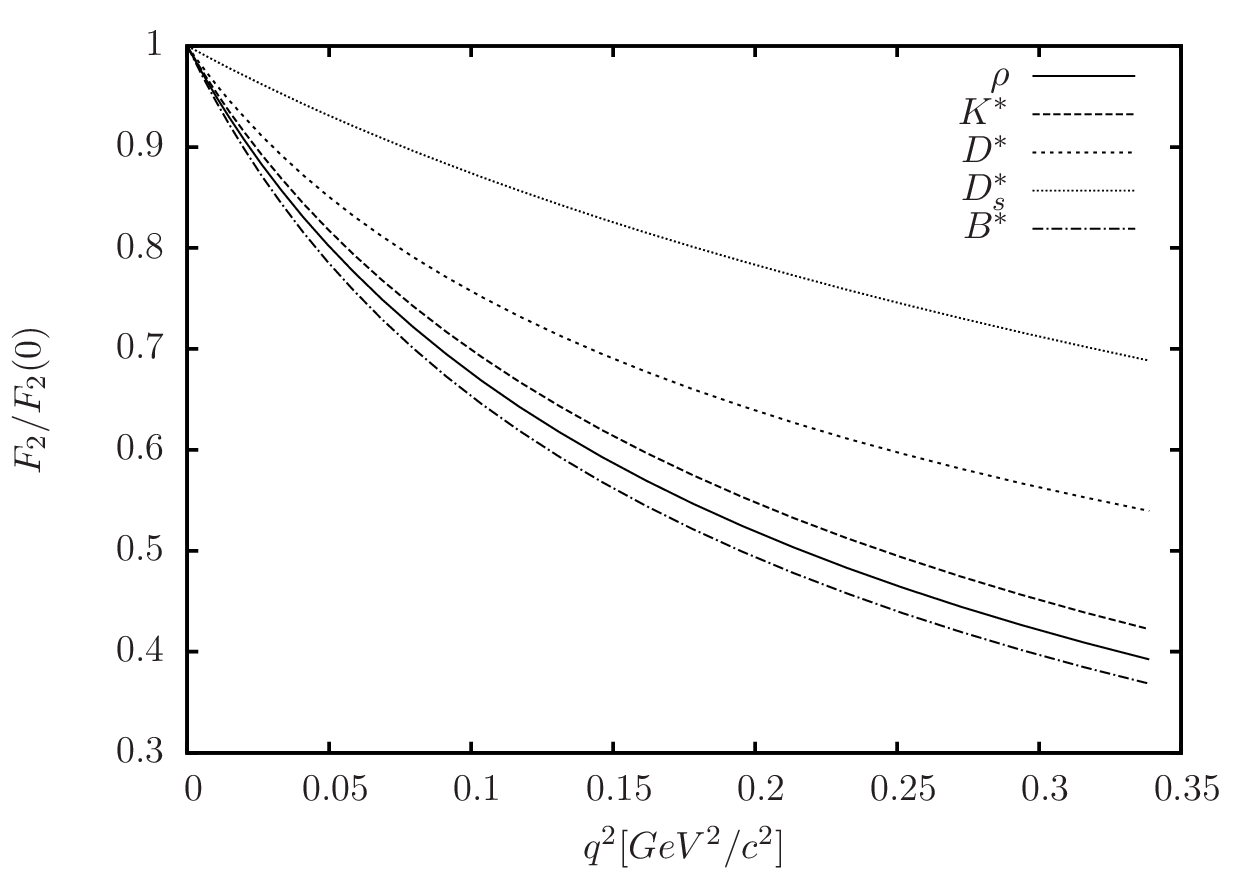}
\caption{$F_2$ of all positive vector mesons with respect to $q^2$}\label{fig9}
\end{center}
\end{figure}

The magnetic form factors $F_2$ are presented in
Fig.~\ref{fig9}. They are connected to the magnetic momentum through
\cite{Hedditch:2007ex}
\begin{align}
\mu_V=\frac{F_2(0)e}{2m_V}.
\end{align}
The magnetic moments are also listed in Table~\ref{tab4}.  The
magnetic moments are given in the unit of nuclear magneton
$\mu_n$. Generally, the magnetic momentum decreases as the meson mass
increases. In our results, the magnetic moments of $D^{\ast}$ and
$D^{\ast}_s$ are smaller than those of the light mesons $\rho$ and
$K$. However, the magnetic moment of $B^{\ast}$ is larger than that of
$D^{\ast}$ and $D^{\ast}_s$. The reason is still that the main
contribution comes from the light quark but the charge of the $u$ is
larger than that of the $d$ and $s$ by a factor of $2$. Up to now, no
experimental data is available. We compare our results for the
$\rho^+$ and $K^{*+}$ mesons with other theoretical work
\cite{Bakker:2002mt} and \cite{Lee2008}.

\begin{table}
\caption{Radii and magnetic moments of vector mesons}\label{tab4}
\begin{ruledtabular}
\begin{tabular}{ccccccccc}
meson&$r^2_1(fm^2)$&$r^2_2(fm^2)$&$r_c(fm)$&$\mu_1(\mu_n)$&$\mu_2(\mu_n)$&$\mu(\mu_n)$&$\mu(\mu_n)$\cite{Bakker:2002mt}&$\mu(\mu_n)$\cite{Lee2008}\\
\hline
$\rho^+$&1.267&1.267&1.12&1.69&0.85&2.54&2.56&3.25\\
$K^{\ast+}$&1.304&0.697&1.05&1.63&0.63&2.26&&2.81\\
$D^{\ast+}$&0.095&1.366&0.72&0.42&0.74&1.16&&\\
$D_s^{\ast+}$&0.083&0.567&0.49&0.42&0.56&0.98&&\\
$B^{\ast+}$&1.359&0.025&0.96&1.4&0.07&1.47&&\\
\end{tabular}
\end{ruledtabular}
\end{table}

\section{SUMMARY}

With the extended NJL model including heavy flavors, we have made a
systematic calculation of the form factors of mesons, including the
pseudo-scalar mesons and their vector partners, of both the light
flavor sector and the light-heavy flavor sector. The form factors of the
$\pi$ and $K$ mesons fit the experimental data.  Other form factors of
mesons, especially of the light-heavy mesons, are presented here to
compare with future experiments and other theoretical calculations such as
lattice calculation.

\begin{acknowledgments}
  We would like to thank Professor Shi-Lin Zhu for useful discussions.
\end{acknowledgments}

\bibliography{form.bib}

\end{document}